\documentclass[a4paper]{article}
\usepackage{geometry}
\usepackage{amsmath}
\usepackage{CJK}
\usepackage{chemarrow}
\usepackage{setspace}
\usepackage{array}
\usepackage{graphicx}
\usepackage{subfigure}
\begin{document}
\begin{CJK*}{GBK}{song}
\begin{spacing}{1.5}
\setlength{\parindent}{2em}
\numberwithin{equation}{section}

\title{Partial equilibrium approximations in Apoptosis\\
%I. The intracellular-signaling subsystem
%Using the principle of detailed balance to simplify apoptosis models
}
\author{Ya-Jing Huang\footnote{Zhou Pei-Yuan Center for Appl. Math., Tsinghua university, Beijing 100084, China; Email: huangyj09@mails.tsinghua.edu.cn},\qquad Wen-An Yong\footnote{Zhou Pei-Yuan Center for Appl. Math., Tsinghua university, Beijing 100084, China; Email: wayong@tsinghua.edu.cn}}
\date{}
\maketitle

 \begin{abstract}
%\boldmath
Apoptosis is one of the most basic biological processes. In apoptosis, tens of species are involved in many biochemical reactions with times scales of widely differing orders of magnitude. By the law of mass action, the process is mathematically described with a large and stiff system of ODEs (ordinary differential equations). The goal of this work is to simplify such systems of ODEs with the PEA (partial equilibrium approximation) method. In doing so, we propose a general framework of the PEA method together with some conditions, under which the PEA method can be justified rigorously. The main condition is the principle of detailed balance for fast reactions as a whole. With the justified method as a tool, we made many attempts via numerical tests to simplify the Fas-signaling pathway model due to Hua et al. (2005) and found that nine of reactions therein can be well regarded as relatively fast. This paper reports our simplification of Hua at el.'s model with the PEA method based on the fastness of the nine reactions, together with numerical results which confirm the reliability of our simplified model.\\

\end{abstract}

\section{Introduction}

Apoptosis is one of the most basic biological phenomena. It is a cellular suicide route that allows for the selective removal of superfluous and potentially dangerous cells. This genetically controlled process ensures normal embryonic development, tissue homeostasis and normal immune-system function in multicellular organisms. On the other hand, defects in apoptosis may cause serious diseases such as cancer, autoimmunity, and neurodegeneration \cite{Hengartner, Horvitz, Thompson, PK}. For these reasons, understanding the mechanism of apoptosis is of fundamental importance.

The apoptotic process involves tens of biological molecules (species), which react within tens of biochemical reactions with time scales of widely differing orders of magnitude. When the law of mass action \cite{Keener} is employed, it is described mathematically by a simultaneous system of tens of ordinary differential equations (ODEs).
Such a large scale and stiff system of ODEs can hardly help us to understand the mechanism of the apoptosis.
%In order to gain insight into the process, it is important to derive a simple but reliable description
%identify their underlying mathematical structure, and to express their dynamics in the most compact way.

The goal of this work is to derive mathematically reliable simplifications of the large apoptosis system proposed by Hua et al. in \cite{Hua} for human Jurkat T cells. Two widely used methods for simplifying chemical kinetics are the Quasi Steady-State Approximation (QSSA) \cite{Benson, Bowen}, also called the Bodenstein method, and Partial Equilibrium Approximation (PEA) \cite{Ramshaw, Rein, Goussis}.
The former assumes that the concentrations of transient intermediate species reach steady states
%\cite{Briggs}. stop to change after a very short time
%Setting equal the rates of production and destruction of these intermediate species
and thereby the rate equations for the intermediate species are replaced with algebraic relations. On the other hand, the PEA simply takes the fast reactions in equilibrium. In this manner, the stiffness is removed and some algebraic constraints are obtained. For both methods, the algebraic relations can be used to reduce the number of the rate equations and consequently the chemical kinetics is simplified. An unexpected benefit of such simplifications is that less parameters are needed for the simplified models than for the original ones. This is good because the parameters are often not reliably known (see \cite{Ramshaw}).

The QSSA, PEA and their combinations have been extensively used to simplify chemical kinetics mechanisms for
many years, with great success \cite{Schott, Miller, Pope, PW, Pe, PR, Smooke}.
%The PEA has been utilized in flame computations for many years \cite{6-12}. Combinations of the PEA and QSSA have
%also been successfully used \cite{13-14}.
%Although the two methods have been used to simplify chemical kinetics mechanisms for many years
%\cite{Keener,Peters,Smooke,Lin},
They were also used by Okazaki et al. in \cite{Okazaki} to simplify the large apoptosis system in \cite{Hua} (see comments below). However, these methods seem to lack a systematically mathematical justification. Recently, we
pointed out that the PEA method can be rigorously justified for reversible reactions obeying the principle of detailed balance \cite{Walls, Yong1}, by using the singular perturbation theory of initial-value problems for ODEs \cite{Yong2}. Thus, our simplification will base on this justified PEA method and therefore is reliable.

As commented in \cite{HY}, Okazaki et al.'s simplification seems baseless. In fact,
when applying the QSSA method for the intermediate species Casp8$_2^*$:Casp3, Okazaki et al. assumed
that the concentration sum of the intermediate species and Casp8$_2^*$ (an activated initiator caspase) was conserved and obtained the
Michaelis-Menten equation for the product (see Appendix A of \cite{Okazaki}). The latter is only true if the activated initiator caspase does not
participate in other reactions. However, this is not the case here because the activated caspase is simultaneously, instead of consecutively, involved
in other reactions. %Namely, the reactions of generating tBid were considered in \cite{Okazaki}
%to be independent of those for the activation of Casp3 by Casp8$_2^*$.
Similar conservation assumptions were used for several steps. This is why we question the
simplified model in \cite{Okazaki}, although it is successful in some sense.

This paper is a continuation of our previous work \cite{HY}. In \cite{HY}, we showed that two molecules (Smac and XIAP) involved in the apoptotic process, neglected in \cite{Okazaki}, are not negligible in general.
Then we applied the justified PEA method to obtain a very preliminary simplified model by assuming only six
reversible reactions to be fast. In the present paper, we will use the PEA method to further simplify the large apoptosis system \cite{Hua}.
To do this, we firstly verify the principle of detailed balance for more fast reactions as a whole.
%At this point, let us remark that the biological system is not closed but open.
Having such a verification, the singular perturbation theory of initial-value problems for ODEs can be employed to derive our simplified models. Then we use numerical simulations to compare the new models with the original one and Okazaki et al.'s simplified model from various aspects, including accuracy, sensitivity and the M-D transition behavior \cite{Okazaki}. Moreover, we introduce a new quantity to evaluate our simplifications. All numerical results confirm the reliability of both our simplified models and the PEA method.

Let us remark that, thanks to its reliability, the justified PEA could be used as a tool to determine whether or not a reversible reaction is relatively fast. To this end, one could numerically compare the solution of partial equilibrium computation with that of the fully non-equilibrium computation. If the two solutions are close to each other, the reaction can be claimed to be relatively fast.

%which determines whether a reaction is assumed to be an equilibrium or finite rate reactions.

The paper is organized as follows. In Section 2 we present a general framework of the PEA method, together with some conditions under which the method can be justified rigorously. The apoptosis process is introduced in Section 3.
The PEA method is used in Section 4 to simplify the apoptosis system by checking the principle of detailed balance
for nine reversible reactions as a whole. Numerical simulations are reported in Section 5. Finally, the main results of this paper are summarized in Section 5.

\section{The PEA Method}

In this section we first present a general framework of the PEA method, together with some conditions under which the method can be justified rigorously. Then we take the simplest system for enzyme inhibition as an example to show how to use our PEA method.

\subsection{A general framework of the PEA method}

Consider a system with $N$ chemical species $C_i (i = 1, 2, \cdots, N)$ participating in $M$ reactions
\begin{equation}\label{21}
a_1^p C_1  + a_2^p C_2  +  \cdots  + a_N^p C_N   \autorightleftharpoons{$k_{+p}$}{$k_{-p}$}  b_1^p C_1  + b_2^p C_2  +  \cdots  + b_N^p C_N
\end{equation}
for $p = 1,2, \cdots ,M$. Here the non-negative integers $a_i^p$ and $b_i^p$ are the stoichiometric coefficients of the $i^{th}$-species in the $p^{th}$-reaction, and $k_{+p}$ and $k_{-p}$ are the respective forward and backward rate constants of
the $p^{th}$-reaction. The reversibility means that both $k_{+p}$ and $k_{-p}$ are positive.

%If $a_i^p  = b_i^p  \ne 0$, then the $p$-th reaction is a catalyzed reaction and $C_i$ is the catalyst for the reaction.
Denote by $u_i=[C_i](t)$ the concentration of the $i^{th}$-species $C_1$ at time $t$. According to the law of mass action \cite{Keener}, the evolution equation for $u_i$ is
\begin{equation}\label{22}
\frac{du_i}{dt} = \sum\limits_{p = 1}^M (b_i^p  - a_i^p )v_p
\end{equation}
with
$$
v_p= {k_{+p} u_1^{a_1^p } u_2^{a_2^p }  \cdots u_N^{a_N^p }  - k_{-p} u_1^{b_1^p } u_2^{b_2^p }  \cdots u_N^{b_N^p } }$$
being the reaction rate of the $p$-th reaction.

Suppose the first $M' (\leq M)$ reactions in (\ref{21}) are much faster than others. Then the kinetic equations in (\ref{22}) can be rewritten in the vectorial form:
\begin{equation}\label{23}
\begin{array}{l}
  \displaystyle{ \frac{{dV}}{{dt}} = \frac{1}{\varepsilon} Q_1(V) + Q_2(V,Z)}, \\[0.2cm]
    \displaystyle{ \frac{{dZ}}{{dt}} = Q_{3}(V,Z)}.
 \end{array}
 \end{equation}
 Here $V$ is an $N'$-vector consisting of those $u_i$ so that the $i^{th}$-species participates in the fast reactions, $Z$ consists of the rest $u_i$, $\varepsilon$ is a small positive parameter characterizing the fastness, and $Q_1(V), Q_2(V, Z)$ and $Q_{3}(V,Z)$ stand for the corresponding reaction rates. The introduction of $\varepsilon$ is to make $Q_1(V)$ have the same order of magnitude as $Q_2(V,Z)$ and $Q_3(V,Z)$. A special case is that $Z$ is void and $V$ contains all $u_i$.

Assume that there is a steady state $u^* =(u_1^*, u_2^*, \cdots u_N^*)$ satisfying $u_i^*>0$ for all $i$ and a zero net flux condition for each fast reaction:
$$
k_{+p} (u_1^* )^{a_1^p } (u_2^* )^{a_2^p }  \cdots (u_N^* )^{a_N^p }  - k_{-p} (u_1^* )^{b_1^p } (u_2^* )^{b_1^p }  \cdots (u_N^* )^{b_1^p }  = 0, \qquad \forall p = 1, 2, \cdots, M'.
$$
This is just the principle of detailed balance \cite{Walls} for the partial system consisting of the fast reactions merely. Clearly, it can only be true when both $k_{+p}$ and $k_{-p}$ are positive, that is, reversible reactions.

Under this assumption, we know from \cite{Yong1} that there is a strictly convex function $\eta=\eta(V)$ so that $Q_1(V)$ can be written as
$$
Q_1(V) = S(V)\eta_V(V)
$$
for $V$ with strictly positive components. Here $S(V)$ is a symmetric matrix with null-space independent of $V$ and
$\eta_V(V)$ is the gradient of $\eta(V)$. Moreover, the singular perturbation theory \cite{Yong2} for initial-value problems of ODEs can be applied to the stiff system in (\ref{23}). In particular, the solutions to initial-value problems of (\ref{23}) converge uniformly to those of a corresponding reduced system, as $\varepsilon$ goes to zero, in any bounded time interval away from zero.

In order to derive the reduced system, we notice the $V$-independence of the null-space and denote by $\Pi$ the constant matrix whose rows span the left null-space of $S(V)$. Without loss of generality, we assume that $\Pi$ is of the form
$$
\Pi = (\Theta, I)
$$
with $I$ the unit matrix of proper order. Accordingly, we introduce the partition
$$
V = \left( {\begin{array}{l}
   X \\
   Y
\end{array}} \right), \qquad Q_1(V) = \left( {\begin{array}{l}
   \hat Q_1(X, Y) \\
   \hat Q_2(X, Y)
\end{array}} \right), \qquad Q_2(V, Z) = \left( {\begin{array}{l}
   \tilde Q_1(X, Y, Z) \\
   \tilde Q_2(X, Y, Z)
\end{array}} \right).
$$
This partition ensures that
$X$ can be uniquely and globally
obtained by solving $\hat Q_1(X, \tilde Y - \Theta X)= 0$ (see \cite{Yong1} if necessary).

Define
$$
\tilde Y = Y + \Theta X .
$$
The kinetic equations in (\ref{23}) can be rewritten as
\begin{equation}\label{24}
\begin{array}{l}
  \displaystyle{ \frac{{dX}}{{dt}} = \frac{1}{\varepsilon}\hat Q_1(X,\tilde Y - \Theta X)  + \tilde Q_1(X,\tilde Y - \Theta X,Z),} \\[0.2cm]
  \displaystyle{ \frac{{d\tilde Y}}{{dt}} = \tilde Q_2(X, \tilde Y - \Theta X, Z) + \Theta \tilde Q_1(X, \tilde Y - \Theta X, Z)} ,\\[0.2cm]
    \displaystyle{ \frac{{dZ}}{{dt}} = Q_3(X, \tilde Y - \Theta X, Z)}.
 \end{array}
 \end{equation}
As $\varepsilon$ goes to zero, the reduced system for (\ref{24}) is
\begin{equation}\label{25}
\begin{array}{l}
  \hat Q_1(X, \tilde Y - \Theta X)  =0,  \\[0.2cm]
  \displaystyle{ \frac{{d\tilde Y}}{{dt}} = \tilde Q_2(X, \tilde Y - \Theta X, Z) + \Theta \tilde Q_1(X, \tilde Y - \Theta X, Z)} ,\\[0.2cm]
    \displaystyle{ \frac{{dZ}}{{dt}} = Q_3(X, \tilde Y - \Theta X, Z)}.
 \end{array}
 \end{equation}
From $\hat Q_1(X, \tilde Y - \Theta X)=0$ we solve $X$ in terms of $\tilde Y$, say $X= \Phi(\tilde Y)$. Substituting this expression into the second and third equations in (\ref{25}), we obtain
\[\left\{ \begin{array}{l}
  \displaystyle{ \frac{{d\tilde Y}}{{dt}} = \tilde Q_2(\Phi(\tilde Y), \tilde Y - \Theta\Phi(\tilde Y), Z)+ \Theta\tilde Q_1(\Phi (\tilde Y), \tilde Y - \Theta\Phi(\tilde Y), Z)}, \\[0.2cm]
    \displaystyle{ \frac{{dZ}}{{dt}} = Q_{3}(\Phi (\tilde Y), \tilde Y - \Theta\Phi(\tilde Y), Z)} .\\
 \end{array} \right.\]
This is our simplified model by using the PEA method. Note that the original variables $X$ and $Y$ are recovered from
$$
X= \Phi(\tilde Y), \qquad Y = \tilde Y - \Theta\Phi(\tilde Y).
$$

In case $X$ can be solved from $\hat Q_1(X, Y)=0$ in terms of $Y$, say $X= \Psi(Y)$, we recall the fact that $X$ can be obtained by solving $\hat Q_1(X, \tilde Y - \Theta X)= 0$ and may well assume that both the Jacobian matrices $\hat Q_{1X}$ (of $\hat Q_1(X, Y)$ with respect to $X$) and $[\hat Q_{1X} - \hat Q_{1Y}\Theta]$ are invertible. Then $[I - \hat Q_{1X}^{-1}\hat Q_{1Y}\Theta]=\hat Q_{1X}^{-1}[\hat Q_{1X} - \hat Q_{1Y}\Theta]$ is invertible. It is an easy exercise to show that the invertibility of $[I - \hat Q_{1X}^{-1}\hat Q_{1Y}\Theta]$ is equivalent to that of $[I - \Theta\hat Q_{1X}^{-1}\hat Q_{1Y}]$. On the other hand, we deduce from $\hat Q_1(\Psi(Y), Y)=0$ that $\hat Q_{1X}\Psi(Y)_Y + \hat Q_{1Y} = 0$ and thereby
$\Psi(Y)_Y=-\hat Q_{1X}^{-1}\hat Q_{1Y}$. Now we compute from $\tilde Y = Y + \Theta\Psi(Y)$ that
$$
\displaystyle{\frac{d\tilde Y}{dt}=\frac{{dY}}{dt}+\Theta\Psi(Y)_{Y}\frac{dY}{dt} } =
[I - \Theta\hat Q_{1X}^{-1}\hat Q_{1Y}]\frac{{dY}}{dt}.
$$
Thus we gain equations for $Y$:
$$
\displaystyle{ \frac{{dY}}{{dt}}=(I + \Theta\Psi(Y)_{Y})^{-1}(\tilde Q_2(\Psi(Y), Y, Z)+ \Theta\tilde Q_1(\Psi(Y), Y, Z))}.
$$
Consequently, the reduced system can be written as
\begin{equation}\label{26}
\begin{array}{l}
  \displaystyle{ \frac{{dY}}{{dt}}=(I +\Theta\Psi(Y)_Y^{-1})( \tilde Q_2(\Psi(Y),Y,Z) + \Theta \tilde Q_1(\Psi(Y),Y,Z))}, \\[0.2cm]
    \displaystyle{ \frac{{dZ}}{{dt}} = Q_{3}(\Psi(Y),Y,Z)}
\end{array}
\end{equation}
together with the algebraic relation $X = \Psi(Y)$.

\subsection{A simple example}

\begin{figure}%[h]
\begin{center}
\includegraphics[width=0.3\textwidth]{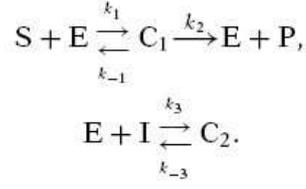}
  \vspace{-.5em}
 \caption{The simplest mechanism for enzyme inhibition}
 \label{Fig:The enzyme system.}
  \vspace{1.2em}
\end{center}
\end{figure}

In order to elucidate how to use the PEA method, we consider the simplest system for enzyme inhibition \cite{Keener}. This system is demonstrated graphically in Fig. \ref{Fig:The enzyme system.} and reveals the competitively inhibitory mechanism, where the enzyme reaction is stopped when the inhibitor is bound to the active site of the enzyme.
In Fig. \ref{Fig:The enzyme system.}, the symbols E, S, I, P, $C_1$ and $C_2$ stand for the enzyme, the substrate, the inhibitor, the product and two complexes, respectively. $k_1$ and $k_{-1}$ are the respective kinetic rate
constants of the forward and backward reaction for substrate binding, while $k_3$ and $k_{-3}$ are
those of the inhibitor binding reaction. $k_2$ is the rate constant of the substrate conversion reaction.

According to law of of mass action, the corresponding kinetic equations read as
\begin{equation}\label{27}
 \begin{array}{l}
 \displaystyle{ \frac{{d[C_1]}}{{dt}} = v_1 - v_2 }  \\[0.2cm]
  \displaystyle{ \frac{{d[C_2]}}{{dt}} = v_3  } \\[0.2cm]
  \displaystyle{ \frac{{d[E]}}{{dt}} = -v_1 - v_3 + v_2 } \\[0.2cm]
  \displaystyle{ \frac{{d[S]}}{{dt}} = -v_1 } \\[0.2cm]
  \displaystyle{ \frac{{d[I]}}{{dt}} = -v_3  } \\[0.2cm]
  \displaystyle{ \frac{{d[P]}}{{dt}} = v_2 }.
 \end{array}
 \end{equation}
with
\[
\begin{array}{l}
 v_1  = k_1[E][S] - k_{-1} [C_1], \\[0.15cm]
 v_2  = k_2[C_1], \\[0.15cm]
 v_3  = k_3[E][I] - k_{-3} [C_2] . \\[0.15cm]
 \end{array}
\]

Classically, the reactions for enzyme to bind to substrates and inhibitors are regarded as fast and reversible. Thus we rewrite (\ref{27}) as
\begin{equation}\label{28}
 \begin{array}{l}
 \displaystyle{ \frac{{d[C_1]}}{{dt}} = \frac{1}{\varepsilon} \hat v_1 - v_2 }  \\[0.2cm]
  \displaystyle{ \frac{{d[C_2]}}{{dt}} = \frac{1}{\varepsilon} \hat v_3  } \\[0.2cm]
  \displaystyle{ \frac{{d[E]}}{{dt}} = -\frac{1}{\varepsilon} \hat v_1 - \frac{1}{\varepsilon}\hat v_3 + v_2 } \\[0.2cm]
  \displaystyle{ \frac{{d[S]}}{{dt}} = -\frac{1}{\varepsilon}\hat v_1 } \\[0.2cm]
  \displaystyle{ \frac{{d[I]}}{{dt}} = -\frac{1}{\varepsilon}\hat v_3  } \\[0.2cm]
  \displaystyle{ \frac{{d[P]}}{{dt}} = v_2 }.
 \end{array}
 \end{equation}
where $\hat v_1 =\varepsilon v_1$, $\hat v_3 =\varepsilon v_3$, and $\hat v_1$ and $\hat v_3$ have the same order of magnitude as $v_2$. Thanks to the reversibility, it is obvious that there are positive numbers $[E]^*, [S]^*, [C_1]^*, [I]^*$ and $[C_2]^*$ such that
$$
v_1 = k_1[E]^*[S]^* - k_{-1} [C_1]^*=0, \qquad v_3 =k_3[E]^*[I]^* - k_{-3} [C_2]^*=0.
$$
Thus the PEA method above can be well applied to the stiff system (\ref{28}).

The reduced system for (\ref{28}) can be derived as follows. Set
$$
\begin{array}{ll}
X =& \big([C_1], [C_2] \big)^T ,\\
Y =& \big([E],[S],[I]\big)^T ,\\
Z =& [P] .
\end{array}
$$
the stiff system (\ref{28}) can be rewritten as
\begin{equation}\label{29}
 \begin{array}{l}
  \displaystyle{ \frac{{dX}}{{dt}} = \frac{1}{\varepsilon}\left( {\begin{array}{*{20}l}
   \hat v_1  \\
   \hat v_3   \\
\end{array}} \right) + v_2\left( {\begin{array}{*{20}l}
   -1\\
   0\\
\end{array}} \right), } \\[0.2cm]
  \displaystyle{ \frac{{d(Y+ \Theta X)}}{{dt}} = v_2\left( {\begin{array}{*{20}l}
  0\\
  -1  \\
   0   \\
\end{array}} \right),  } \\[0.2cm]
  \displaystyle{ \frac{{d[P]}}{{dt}} = v_2 }
 \end{array}
 \end{equation}
with
$$
\Theta =\left( {\begin{array}{*{20}l}
   {1} & {1}  \\
   {1} & {0}  \\
   {0} & 1   \\
\end{array}} \right).
$$
From $\hat v_1=\hat v_3=0$, we get
\begin{equation}\label{210}
[C_1] = K_1[E][S] , \qquad [C_2] = K_3[E][I]
\end{equation}
with $K_j=k_j/k_{-j}$ for $j= 1, 3$. Substituting these into the last two equations in (\ref{29}), we obtain the following ODEs
$$
 \begin{array}{l}
  \displaystyle{ \frac{{d([E]+[C_1]+[C_2])}}{{dt}} = 0, } \\[0.2cm]
  \displaystyle{ \frac{{d([S]+[C_1])}}{{dt}} = -k_2 K_1 [E][S],  } \\[0.2cm]
  \displaystyle{ \frac{{d([I]+[C_2]) }}{{dt}} = 0,  } \\[0.2cm]
  \displaystyle{ \frac{{d[P]}}{{dt}} = k_{2} K_1 [E][S] }.
 \end{array}
$$
This, together with (\ref{210}), is the simplified model by using the justified PEA method. This model can be further simplified by using the conservation laws indicated in the first and third equations. We omit it here and leave it to the interested reader.

\section{Apoptosis Systems}

Here we introduce the large apoptosis system proposed by Hua et al. in \cite{Hua} for human Jurkat T cells. To begin with, we recall that there are at least two pathways to trigger apoptosis---intrinsic (mitochondrial) and extrinsic (death receptor) signalling pathways. Both induce death-associated proteolytic and/or nucleolytic activities. The intrinsic pathway is initiated when the cell is severely damaged or stressed, while the extrinsic one is activated when extracellular death ligands are bound by their cognate membrane-associated death receptors such as TNF-R1(DR1,p55), Fas(DR2,CD95), DR3(APO-3,TRAMP), DR4(APO-2,TRAIL-R1) and DR5(TRICK2,TRAOL-R2)
\cite{Hengartner,Ashkenazi,Barnhart,Samraj,Lavrik}.
%,Schmitz}.

\begin{figure}%[h]
  \begin{center}
  \includegraphics[width=1\textwidth]{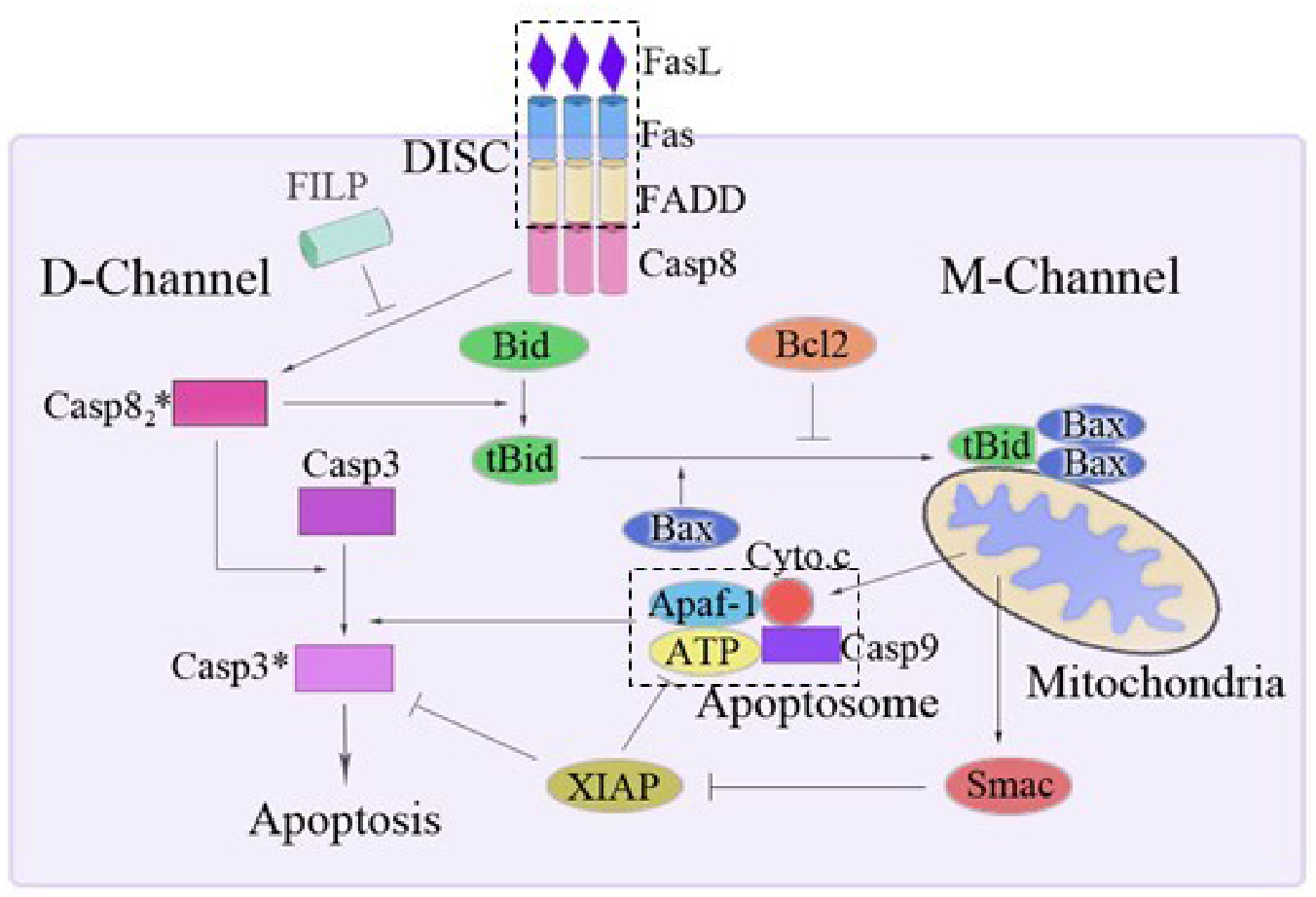}
  %\vspace{-.5em}
  \caption{The Fas-induced apoptotic pathway, including two channels.}
  \label{Fig:The cell apoptosis picture}
  %\vspace{1.2em}
  \end{center}
\end{figure}

The Fas-induced signaling pathway is among the best understood and can be schematically shown in Fig. \ref{Fig:The cell apoptosis picture}.
It begins with the binding of Fas ligands (FasL), Fas and FADD (Fas-associated death domain) to form the
complex DISC (death-inducing signaling complex). The latter can recruits initiator caspases such as caspase-8 (Casp8) molecules to cleave and
activate them. The activated initiator caspase (Casp8$_2^*$) can cleaves and activates the executor caspase-3 (Casp3)
to form Casp3$^*$ directly. The amount of Casp3$^*$ is the indicator of apoptosis. This way to activate
Casp3 is called D-channel. In addition, Casp3 can also be activated in a so-called M-channel. In this channel,
%Casp8$^*$ activates the Casp3 directly in D channel, whereas the process is not such simply for M channel. In the M channel, the role of
Casp8$_2^*$ cleaves Bid to generate truncated (t)Bid.
%but without any change to itself.
The tBid then binds to two molecules of Bax to form a complex tBid:Bax$_2$, which will induce the release of Cyto.c and Smac
from the mitochondria. The released Cyto.c$^*$ will combine an adaptor protein Apaf-1, ATP and caspase-9 to form apoptosome and thereby
activate caspase-9. The activated caspase-9 (Casp9$^*$) cleaves and activates Casp3. On the other hand, the
M-channel can be blocked by XIAP (X-linked inhibitor of apoptosis protein) and Bcl2 through their bindings to
the released Smac*, Casp9, Casp3*, Bax and tBid.
%corresponding action whereby the released Smac* can combine to XIAP to eliminate this effect.
%Once Casp3 is activated, it will trigger a cascade of events leading to apoptosis.

The Fas-signaling pathway model proposed by Hua et at. \cite{Hua} consists of biochemical reactions (H1)--(H25) given in Table 1.
\begin{table}
\tiny
\vspace{0cm}
\caption{The Fas-signaling pathway model due to Hua et at. (2005)}
\medskip{}
\hspace{0cm}
\label{Tab:the ISS model}
\begin{tabular}{llll}
\hline
Reaction & \space & $k_{i}$ & $ k_{-i}$  \\
\hline
(H1) & $ FasL + Fas \autorightleftharpoons{$k_{H1}$}{$k_{-H1}$} FasC $  & $9.09\times10^{-5}nM^{-1}s^{-1}$ & $ 1.00\times10^{-4}$ \\
(H2)$^a$ & $ FasC:FADD_p:Casp8_q:FLIP_r + FADD \autorightleftharpoons{$k_{H2}$}{$k_{-H2}$} Fas:FADD_{p+1}:Casp8_q:FILP_r $ & $5.00\times10^{-4}nM^{-1}s^{-1}$ &  0.2 \\
(H3)$^b$ & $ FasC:FADD_p:Casp8_q:FILP_r + Casp8 \autorightleftharpoons{$k_{H3}$}{$k_{-H3}$} Fas:FADD_p:Casp8_{q+1}:FILP_r $ & $3.50\times10^{-3}nM^{-1}s^{-1}$  & 0.018 \\
(H4)$^b$ & $ FasC:FADD_p:Casp8_q:FLIP_r + FILP \autorightleftharpoons{$k_{H4}$}{$k_{-H4}$} Fas:FADD_p:Casp8_q:FILP_{r+1} $ & $3.50\times10^{-3}nM^{-1}s^{-1}$ & 0.018 \\
(H5)$^c$ & $ FasC:FADD_p:Casp8_q:FLIP_r  \autorightarrow{$k_{H5}$}{} Casp8*_2:p41 + FasC:FADD_p:Casp8_{q-1}:FILP_r $ & $0.3s^{-1}$ \\
(H6) & $ Casp8_{_2 }^*:p41 \autorightarrow{$k_{H6}$}{} Casp8_{_2 }^* $ & $0.1s^{-1}$ & \\
(H7) & $Casp8_{_2 }^*  + Casp3  \autorightleftharpoons{$k_{H7}$}{$k_{-H7}$}   Casp8_{_2 }^* :Casp3 $ & $1.00\times10^{-4}nM^{-1}s^{-1}$ &  0.06 \\
(H8) & $ Casp8_{_2 }^* :Casp3    \autorightarrow{$k_{H8}$}{}   Casp8_{_2 }^*  + Casp3^* $ & $0.1s^{-1}$ \\
(H9) & $ Casp8_{_2 }^*  + Bid    \autorightleftharpoons{$k_H9$}{$k_{-H9}$}   Casp8_{_2 }^* :Bid  $ & $5.00\times10^{-4}nM^{-1}s^{-1}$ &  0.005 \\
(H10) & $ Casp8_{_2 }^* :Bid      \autorightarrow{$k_{H10}$}{}   Casp8_{_2 }^*  + tBid $ & $0.1s^{-1}$  \\
(H11) & $ tBid + Bax        \autorightleftharpoons{$k_{H11}$}{$k_{{-H11}}$}  tBid:Bax  $ & $2.00\times10^{-4}nM^{-1}s^{-1}$ &  0.02 \\
(H12) & $ tBid:Bax + Bax    \autorightleftharpoons{$k_{H12}$}{$k_{-H12}$}  tBid:Bax_2 $ & $2.00\times10^{-4}nM^{-1}s^{-1}$ &  0.02 \\
(H13) & $ Smac + tBid:Bax_2 \autorightarrow{$k_{H13}$}{} Smac^* + tBid:Bax_2 $ & $1.00\times10^{-3}nM^{-1}s^{-1}$ \\
(H14) & $ Smac^* + XIAP \autorightleftharpoons{$k_{H14}$}{$k_{{-H14}}$} Smac^*:XIAP $ & $7.00\times10^{-3}nM^{-1}s^{-1}$  & $2.21\times10^{-3}$ \\
(H15) & $ Cyto.c + tBid:Bax_2   \autorightarrow{$k_{H15}$}{}  Cyto.c^*  + tBid:Bax_2  $ & $1.00\times10^{-3}nM^{-1}s^{-1}$  \\
(H16) & $ Cyto.c^* + Apaf + ATP \autorightleftharpoons{$k_{H16}$}{$k_{-H16}$}   Cyto.c^* :Apaf:ATP $ & $2.78\times10^{-7}nM^{-1}s^{-1}$ &  $5.70\times10^{-3}$ \\
(H17) & $ Cyto.c^* :Apaf:ATP + Casp9 \autorightleftharpoons{$k_{H17}$}{$k_{-H17}$}  Cyto.c^* :Apaf:ATP:Casp9  $ & $2.84\times10^{-4}nM^{-1}s^{-1}$ &  0.07493 \\
(H18) & $ Cyto.c^* :Apaf:ATP:Casp9 + Casp9 \autorightleftharpoons{$k_{H18}$}{$k_{-H18}$}  Cyto.c^* :Apaf:ATP:Casp9_2 $ & $4.41\times10^{-4}nM^{-1}s^{-1}$ &  0.1 \\
(H19) & $ Cyto.c^* :Apaf:ATP:Casp9_2 \autorightarrow{$k_{H19}$}{}  Cyto.c^* :Apaf:ATP:Casp9 + Casp9^* $ & $0.7s^{-1}$ \\
(H20) & $ Casp9^*  + Casp3  \autorightleftharpoons{$k_{H20}$}{$k_{-H20}$}  Casp9^* :Casp3 $ & $1.96\times10^{-5}nM^{-1}s^{-1}$ &  0.05707 \\
(H21) & $ Casp9^* :Casp3    \autorightarrow{$k_{H21}$}{}   Casp9^*  + Casp3^* $ & $4.8s^{-1}$ \\
(H22) & $ Casp9 + XIAP \autorightleftharpoons{$k_{H22}$}{$k_{{-H22}}$} Casp9:XIAP $ & $1.06\times10^{-4}nM^{-1}s^{-1}$ &  $1.00\times10^{-3}$ \\
(H23) & $ Casp3^* + XIAP \autorightleftharpoons{$k_{H22}$}{$k_{{-H22}}$} Casp3^*:XIAP $ & $2.47\times10^{-3}nM^{-1}s^{-1}$ &  $2.40\times10^{-3}$  \\
(H24) & $ Bcl_2  + Bax      \autorightleftharpoons{$k_{H24}$}{$k_{-H24}$}  Bcl_2 :Bax $ & $2.00\times10^{-4}nM^{-1}s^{-1}$ &  0.02 \\
(H25) & $ Bcl_2  + tBid     \autorightleftharpoons{$k_{H25}$}{$k_{-H25}$}  Bcl_2 :tBid $ & $2.00\times10^{-4}nM^{-1}s^{-1}$ &  0.02 \\
\hline
\end{tabular}\\[4mm]
{\small
The index $(p, q, r)$ in reactions (a) takes values (0,0,0),(1,0,0),(1,0,1),(1,1,0),(2,0,0),(2,0,1),(2,0,2),(2,1,0),(2,1,1) and (2,2,0).
In reactions (b) it takes values (1,0,0),(2,0,0),(2,0,1),(2,1,0),(3,0,0),(3,0,1),(3,0,2),(3,1,0),(3,1,1) and (3,2,0), while it takes values (2,2,0),(3,2,0),(3,2,1) and (3,3,0) in reactions (c). }
\end{table}
From this table we see that the process activating the initiator caspase-8 (Casp8) consists of the reactions from (H1) to (H6), which is initiated by FasL. The activated Casp$8_2^*$ enzymatically cleaves caspase-3(Casp3) to produce activated executor Casp$3^*$ ((H7) and (H8)) and Bid to generate tBid ((H9) and (H10)) simultaneously. Then the tBid associates with two Bax to form tBid:Bax$_2$ through (H11) and (H12), which induces the release of Cyto.c and Smac from mitochondrial to cytosol ((H15) and (H13)). The released Cyto.c (Cyto.c$^*$) combines Apaf (Apaf-1) and ATP to form an apoptosome (Cyto.c$^*$:Apaf:ATP) in (H16), which recruits two caspase-9(Casp9) and generates the activated caspase-9 (Casp$9^*$) through (H17) to (H19). The activated Casp$9^*$ can also enzymatically cleaves and activates caspase-3  ((H20) and (H21)). On the other hand, the roles of Casp9, Casp$3^*$, tBid and Bax can be inhibited by binding to XIAP((H22) and (H23)) and Bcl$_2$ ((H24) and (H25)), while the released Smac (Smac*) can suppress the function of XIAP (H14). Table 1 also contains all the forward/backward rate constants $k_{\pm i}(i =1, 2, \cdots, 25)$.

Observe that not every reaction in Table 1 is reversible and the reactions activating Casp8 are independent of the rest. As in \cite{Okazaki}, we call the downstream process, consisting of reactions from (H7) to (H25), as the intracellular-signaling subsystem (ISS). Moreover, we follow \cite{Okazaki} and assume that the concentration of ATP is a fixed constant. Thus, there are 28 species and 19 biochemical reactions involved in the downstream process.

According to the law of mass action, the dynamics of the ISS is governed by 28 ordinary differential equations
\begin{equation}\label{31}
\displaystyle{ \frac{{dU}}{{dt}} = Q(U)}.
\end{equation}
Here $U=U(t)$ is a column vector with 28 components representing the concentrations of all the 28 species in the ISS:
$$
\begin{array}{ll}
U =& \big([Casp8^*_2], [Casp8^*_2:Casp3], [Casp8^*_2:Bid], [Bid], [tBid], [tBid:Bax], [tBid:Bax_2], \\
 &  \ \ [Bcl_2:tBid], [Bax], [Bcl_2:Bax], [Bcl_2], [Cyto.c],  [Cyto.c^*], [Cyto.c^* :Apaf:ATP], \\
 & \ \ [Cyto.c^*:Apaf:ATP:Casp9], [Cyto.c^*:Apaf:ATP:Casp9_2], [Apaf], [Casp9^*],\\
 & \ \ [Casp9], [Casp3],  [Casp9*:Casp3], [Casp3^*], [Smac], [Smac^*], [XIAP], \\
 & \ \ [Smac^*:XIAP], [Casp9:XIAP], [Casp3^*:XIAP] \big)^T ,
\end{array}
$$
each element of the vector-valued function $Q(U)$ of $U$ is the change rate of concentration for the corresponding species
$$
\begin{array}{ll}
Q(U) = & \big(  -v_{7} + v_{8} - v_{9} + v_{10} + v_0, v_{7}-v_{8}, v_{9}-v_{10}, -v_{9}, v_{10} - v_{11}- v_{25}, v_{11}-v_{12}, v_{12}, v_{25}, \\
& \ \  -v_{11}-v_{12}-v_{24},
v_{24}, -v_{24}-v_{25}, -v_{15}, v_{15}-v_{16},  v_{16}-v_{17}, v_{17}-v_{18}+v_{19}, \\ & \ \  v_{18}-v_{19}, -v_{16}, v_{19}-v_{20}+v_{21},
-v_{17}-v_{18}-v_{22}, -v_{7}-v_{20}, v_{20}-v_{21},  \\ &  \ \ v_{8}+v_{21}-v_{23}, -v_{13}, v_{13}-v_{14}, -v_{14}-v_{22}-v_{23}, v_{14}, v_{22},v_{23} \big)^T
\end{array}
$$
with $v_i (i=7\cdots25)$ the rate of the $i$-th reaction in Table 1:
\[
\begin{array}{l}
 v_7  = k_7 [Casp8_{_2 }^* ][Casp3] - k_{-7} [Casp8_{_2 }^* :Casp3], \\[0.2cm]
 v_8  = k_{8} [Casp8_{_2 }^* :Casp3], \\[0.2cm]
 v_9  = k_{9} [Casp8_{_2 }^* ][Bid] - k_{ - 9} [Casp8_{_2 }^* :Bid], \\[0.2cm]
 v_{10}  = k_{10} [Casp8_{_2 }^* :Bid], \\[0.2cm]
 v_{11}  = k_{11} [tBid][Bax] - k_{ - 11} [tBid:Bax], \\[0.2cm]
 v_{12}  = k_{12} [tBid:Bax][Bax] - k_{ - 12} [tBid:Bax_2 ], \\[0.2cm]
 v_{13}  = k_{13} [Smac][tBid:Bax_2]  \\[0.2cm]
 v_{14}  = k_{14} [Smac^*][XIAP] - k_{-14} [Smac^*:XIAP] , \\[0.2cm]
 v_{15}  = k_{15} [Cyto.c][tBid:Bax_2 ] ,\\[0.2cm]
 v_{16}  = k_{16} [Cyto.c^* ][Apaf][ATP] - k_{ - 16} [Cyto.c^* :Apaf:ATP] ,\\[0.2cm]
 v_{17}  = k_{17} [Cyto.c^* :Apaf:ATP][Casp9] - k_{ - 17} [Cyto.c^* :Apaf:ATP:Casp9] ,\\[0.2cm]
 v_{18}  = k_{18} [Cyto.c^* :Apaf:ATP:Casp9][Casp9] - k_{ - 18} [Cyto.c^* :Apaf:ATP:Casp9_2 ], \\[0.2cm]
 v_{19}  = k_{19} [Cyto.c^* :Apaf:ATP:Casp9_2 ], \\[0.2cm]
 v_{20}  = k_{20} [Casp9^* ][Casp3] - k_{ - 20} [Casp9^* :Casp3], \\[0.2cm]
 v_{21}  = k_{21} [Casp9^* :Casp3], \\[0.2cm]
 v_{22}  = k_{22} [Casp9][XIAP] - k_{-22} [Casp9:XIAP] , \\[0.2cm]
 v_{23}  = k_{23} [Casp3^*][XIAP] - k_{-23} [Casp3^*:XIAP] , \\[0.2cm]
 v_{24}  = k_{24} [Bcl_2 ][Bax] - k_{ - 24} [Bcl_2 :Bax], \\[0.2cm]
 v_{25}  = k_{25} [Bcl_2 ][tBid] - k_{ - 25} [Bcl_2 :tBid],
 \end{array}
\]
$v_0$ is the constant rate of generation for Casp8$^*_2$ from the upstream process and its value was suggested in \cite{Okazaki} as $v_0=0.001nMs^{-1}$.
%Note that none of the reactions will occur if Casp8$^*_2$ has no a source.
In addition, the non-zero initial concentrations for $U$ are taken as in \cite{Hua, Okazaki} and are given in Table
\ref{Tab:The initial concentrations of each species in ISS model}.
\begin{table}
\tiny
\vspace{0cm}
\caption{Non-zero initial concentrations of the species in the ISS model (Hua. et al. 2005)}
\medskip{}
\hspace{5cm}
\label{Tab:The initial concentrations of each species in ISS model}
\begin{tabular}{lr}
 \hline
Species & Initial concentration(nM)   \\
\hline
 Casp3  & 200.00  \\
 Bid    & 25.00 \\
 Bcl2   & 75.00  \\
 Bax    & 83.33 \\
 Cyto.c & 100.00 \\
 Smac   & 100.00 \\
 XIAP   & 30.00 \\
 Casp9  & 20.00 \\
 ATP    & 10000.00 \\
 Apaf   & 100.00 \\
\hline
\end{tabular}
\end{table}

In \cite{Okazaki}, Okazaki et al. claimed that Smac and XIAP have little effect on the reaction process of the ISS
by studying the M-D transition behavior of both the ISS model and ISS(wo/S,X) (without Smac and XIAP) model.
So they did not consider reactions (H13), (H14), (H22) and (H23) in their simplification. However, in our previous paper \cite{HY} we found that Smac and XIAP should not be ignored because the numerical results of these two models are quite different if initial concentrations are changed. Therefore, our sequel discussion will base on the entire ISS system.

%With the above parameters and initial values, Okazaki et al. \cite{Okazaki} investigated the M-D transition behaviors
%of the ISS and ISS(wo/S,X) (without Smac and XIAP). In doing this, they introduced a quantity $\gamma _D$ to
%characterize the net production rate of Casp3$^*$ by the D-channel. By numerical simulations, they claimed that
%Smac and XIAP have little effect on the reaction process of the ISS and therefore they did not consider the related
%reactions in their simplification. We repeat their numerical results
%(see Fig. \ref{Fig.The comparison of ISS and ISS(wo/S,X)1}) and agree with that their claim is reasonable for the
%above data.

We conclude this section by explaining the M-D transition behavior. It means a D-channel and M-channel switching
behavior and the quantity of [Casp8$_2^*$] is a control parameter. When a large amount of Casp8$_2^*$ is activated
from the upstream process, it will directly induce cell death through the D-channel; otherwise, the M-channel plays
more important role for cell death. In \cite{Hua}, the authors claimed that
%Bcl2 interaction with both Bax and tBid is the most likely mechanism for Bcl2 to block the mitochondrial pathway and
the effects of D-channel and M-channel can be altered by varying the amount of Casp8$_2^*$ generated by DISC
\cite{Barnhart,Samraj,Scaffidi}, which is consistent with previous experiments.

\section{Model reduction}

In this section we use the PEA method to investigate the large apoptosis system (\ref{31}). This system contains 13 reversible reactions: (H7), (H9), (H11), (H12), (H14), (H16), (H17), (H18), (H20), (H22), (H23), (H24) and (H25).
In our preliminary work \cite{HY}, we showed that the six reactions (H11), (H12), (H16), (H17), (H24) and (H25) are
fast. Because the PEA method has a solid mathematical basis, it can be used as a tool to determine whether or not a
reversible reaction is fast. After many attempts by assuming some of the rest 7 reactions to be fast too, we find that the nine reversible reactions (H7), (H9), (H11), (H12), (H16), (H17), (H18), (H24) and (H25) can be well regarded as fast.

%In [30], enzyme reactions (H7), (H8), (H9), (H10), (H18), (H19), (H20) and (H21) were treated by QSSA method meanwhile
%reactions (H13), (H14), (H22) and (H23) were ignored directly.
%Here, we apply the justified PEA method to do the simplification by respectively assuming any subset of reversible
%reactions (H7), (H9), (H18), (H20), (H14), (H22) and (H23) to be fast, besides (H11), (H12), (H16), (H17), (H24)
%(H14), and (H25).

%Through various attempts, we found that the simplification by adding assumption of reversible reactions (H7), (H9)
%and (H18) to be fast is the most reasonable. Any other assumption that contains one reaction of (H18), (H20), (H14),
%(H22) and (H23) to be fast was confirmed to be undesirable, since the related simulation results of these kind of
%simplification are quite different from that of ISS model.

Here we derive the corresponding simplified model under the assumption that the nine reversible reactions are fast. According to the framework in Section 2, we decompose the concentration vector $U$ as
$$
U = \left( {\begin{array}{l}
   X \\
   Y \\
   Z
\end{array}} \right).
$$
Here $X$ stands for the products of the nine reactions, $Y$ for the reactants and $Z$ for the rest:
\begin{equation}\label{41}
\begin{array}{ll}
X =& \big( [Casp8^*_2:Casp3], [Casp8^*_2:Bid], [tBid:Bax], [tBid:Bax_2], \\
& ~ [Bcl_2:Bax], [Bcl_2:tBid], [Cyto.c^* :Apaf:ATP], \\
& ~ [Cyto.c^*:Apaf:ATP:Casp9],[Cyto.c^*:Apaf:ATP:Casp9_2] \big)^T, \\
Y = & \big( [Casp8^*_2], [Bid], [tBid], [Bax], [Bcl_2], [Cyto.c^*], [Apaf], [Casp9], [Casp3] \big)^T, \\
Z = & \big(  [Cyto.c], [Casp9^*] ,  [Casp9*:Casp3],[Casp3^*] , [Smac], [Smac]^*, [XIAP],\\
& ~[Smac^*:XIAP], [Casp9:XIAP], [Casp3^*:XIAP] \big)^T.
\end{array}
\end{equation}
With this decomposition, the kinetic equations in (\ref{31}) can be rewritten as
\begin{equation}\label{42}
 \begin{array}{l}
  \displaystyle{ \frac{{dX}}{{dt}} = \frac{1}{\varepsilon}\hat Q_{1}(X,Y)  + Q_{1}(X,Y,Z)} \\[0.2cm]
  \displaystyle{ \frac{{dY}}{{dt}} = \frac{1}{\varepsilon}\hat Q_{2}(X,Y)  + Q_{2}(X,Y,Z)} \\[0.2cm]
    \displaystyle{ \frac{{dZ}}{{dt}} = Q_{3}(X,Y,Z)}.
 \end{array}.
  \end{equation}
Here the small parameter $\varepsilon$ characterizes the fastness of the reversible reactions as in Section 2,
$$
\begin{array}{ll}
\hat Q_1(X,Y) = & \varepsilon \big( v_7, v_9, v_{11}-v_{12}, v_{12},v_{24},v_{25},v_{16}-v_{17},v_{17}-v_{18},v_{18} \big)^T, \\
\hat Q_2(X,Y) = & \varepsilon \big( -v_7-v_9, -v_9, -v_{11}-v_{25}, -v_{11}-v_{12}-v_{24}, -v_{24}-v_{25}, -v_{16}, -v_{16}, -v_{17}-v_{18},-v_7 \big)^T,
\end{array}
$$
stand for the change rates of concentration due to the rapid reactions, and
$$
\begin{array}{ll}
Q_{1}(X,Y,Z) = & \big( -v_{8}, -v_{10}, 0,0,0,0,0, v_{19}, -v_{19} \big)^T, \\
Q_{2}(X,Y,Z)=& \big( v_{8}+v_{10}+v_0, 0, v_{10}, 0, 0, v_{15}, 0,-v_{22},  -v_{20} \big)^T, \\
Q_{3}(X,Y,Z) = & \big( -v_{15}, v_{19}-v_{20}+v_{21}, v_{20}-v_{21}, v_8 + v_{21}-v_{23} , -v_{13}, v_{13}-v_{14},-v_{14}-v_{22}-v_{23},v_{14}, v_{22},v_{23}\big)^T,
\end{array}
$$
are those for the other reactions. %Note that $\hat Q_1(X,Y)$ is same as $Q_i(X,Y, Z) (i = 1, 2, 3)$ in order of magnitude.
It is direct to check that
\begin{equation}\label{43}
\hat Q_{2}(X,Y) + \Theta\hat Q_{1}(X,Y)\equiv 0
\end{equation}
with $\Theta$ the following constant 9x9-matrix
$$
\Theta =\left( {\begin{array}{*{20}l}
   {1} & {1} & {0} & {0} & {0} & {0} & {0} & {0} & {0}  \\
   {0} & {1} & {0} & {0} & {0} & {0} & {0} & {0} & {0}  \\
   {0} & {0} & {1} & {1} & {0} & {1} & {0} & {0} & {0}  \\
   {0} & {0} & {1} & {2} & {1} & {0} & {0} & {0} & {0}  \\
   {0} & {0} & {0} & {0} & {1} & {1} & {0} & {0} & {0}  \\
   {0} & {0} & {0} & {0} & {0} & {0} & {1} & {1} & {1}  \\
   {0} & {0} & {0} & {0} & {0} & {0} & {1} & {1} & {1}  \\
   {0} & {0} & {0} & {0} & {0} & {0} & {0} & {1} & {2}  \\
   {1} & {0} & {0} & {0} & {0} & {0} & {0} & {0} & {0}  \\
\end{array}} \right).
$$

Recall that
\begin{equation}\label{44}
\begin{array}{l}
 v_7  = k_{7} [Casp8_{_2 }^* ][Casp3] - k_{- 7} [Casp8_{_2 }^* :Casp3], \\[0.1cm]
 v_9  = k_{9} [Casp8_{_2 }^* ][Bid] - k_{-9} [Casp8_{_2 }^* :Bid], \\[0.1cm]
 v_{11}  = k_{11} [tBid][Bax] - k_{-11} [tBid:Bax] , \\[0.1cm]
 v_{12}  = k_{12} [tBid:Bax][Bax] - k_{ -12} [tBid:Bax_2] , \\[0.1cm]
 v_{16}  = k_{16} [Cyto.c^* ][Apaf][ATP] - k_{ -16} [Cyto.c^* :Apaf:ATP] ,\\[0.1cm]
 v_{17}  = k_{17} [Cyto.c^* :Apaf:ATP][Caps9] - k_{ -17} [Cyto.c^* :Apaf:ATP:Casp9] , \\[0.1cm]
 v_{18}  = k_{18} [Cyto.c^* :Apaf:ATP:Casp9][Casp9] - k_{ - 18} [Cyto.c^* :Apaf:ATP:Casp9_2 ], \\[0.1cm]
 v_{24}  = k_{24} [Bcl_2][Bax] - k_{ -24} [Bcl_2 :Bax] , \\[0.1cm]
 v_{25}  = k_{25} [Bcl_2][tBid] - k_{ -25} [Bcl_2 :tBid].
 \end{array}
 \end{equation}
Then for any given $Y = \big( [Casp8^*_2], [Bid], [tBid], [Bax], [Bcl_2], [Cyto.c^*], [Apaf], [Casp9], [Casp3] \big)^T$ with positive components, we use (\ref{43}) to get
$$
\begin{array}{ll}
X =& \big( [Casp8^*_2:Casp3], [Casp8^*_2:Bid], [tBid:Bax], [tBid:Bax_2],  [Bcl_2:Bax], [Bcl_2:tBid], \\
 &~[Cyto.c^* :Apaf:ATP], [Cyto.c^*:Apaf:ATP:Casp9],[Cyto.c^*:Apaf:ATP:Casp9_2] \big)^T
\end{array}
$$
with positive components such that $v_7=v_9=v_{11}=v_{12}=v_{16}=v_{17}=v_{18}=v_{24}=v_{25}=0$. Thus, the principle of detailed balance is verified.

After verifying the conditions for the PEA method to be reliable, we turn to write down the simplified model. In view of (\ref{43}), we define $\tilde Y = Y + \Theta X.$ Then the ODEs in (\ref{42}) become
\begin{equation}\label{45}
\begin{array}{l}
  \displaystyle{ \frac{{dX}}{{dt}} = \frac{1}{\varepsilon}\hat Q_1(X,Y)  + Q_1(X,Y,Z),} \\[0.2cm]
  \displaystyle{ \frac{{d\tilde Y}}{{dt}} = Q_2(X, Y, Z) + \Theta Q_1(X, Y, Z)} ,\\[0.2cm]
    \displaystyle{ \frac{{dZ}}{{dt}} = Q_3(X, Y, Z)}.
 \end{array}
 \end{equation}
Guided by the framework in Section 2, we solve $\hat Q_1(X, Y)=0$, namely,
\[\left\{ \begin{array}{l}
 v_7  = k_{7} [Casp8_{_2 }^* ][Casp3] - k_{ -7} [Casp8_{_2 }^* :Casp3]=0, \\[0.1cm]
 v_9  = k_{9} [Casp8_{_2 }^* ][Bid] - k_{ - 9} [Casp8_{_2 }^* :Bid]=0, \\[0.1cm]
 v_{11}  = k_{11} [tBid][Bax] - k_{ -11} [tBid:Bax]=0 , \\[0.1cm]
 v_{12}  = k_{12} [tBid:Bax][Bax] - k_{ -12} [tBid:Bax_2]=0 , \\[0.1cm]
 v_{16}  = k_{16} [Cyto.c^* ][Apaf][ATP] - k_{ -16} [Cyto.c^* :Apaf:ATP]=0 ,\\[0.1cm]
 v_{17}  = k_{17} [Cyto.c^* :Apaf:ATP][Caps9] - k_{ -17} [Cyto.c^* :Apaf:ATP:Casp9] =0, \\[0.1cm]
 v_{18}  = k_{18} [Cyto.c^* :Apaf:ATP:Casp9][Casp9] - k_{ - 18} [Cyto.c^* :Apaf:ATP:Casp9_2 ]=0, \\[0.1cm]
 v_{24}  = k_{24} [Bcl_2][Bax] - k_{ -24} [Bcl_2 :Bax]=0 , \\[0.1cm]
 v_{25}  = k_{25} [Bcl_2][tBid] - k_{ -25} [Bcl_2 :tBid]=0.
 \end{array} \right.\]
From these algebraic equations we can easily solve $X$ in terms of $Y$:
\begin{equation}\label{46}
\begin{array}{rl}
 [Casp8_{_2 }^* :Casp3] = & \frac{{k_{7} [Casp8_{_2 }^* ][Casp3]}}{{k_{ -7} }} = {K_{7} [Casp8_{_2 }^* ][Casp3]}, \\[0.15cm]
 [Casp8_{_2 }^* :Bid] = & \frac{{k_{9} [Casp8_{_2 }^* ][Bid]}}{{k_{ - 9} }} = {K_{9} [Casp8_{_2 }^* ][Bid]}, \\[0.15cm]
 [tBid:Bax] = & \frac{{k_{11} [tBid][Bax]}}{{k_{ -11} }} = K_{11} [tBid][Bax], \\[0.15cm]
 [tBid:Bax_2] =  & \frac{{k_{12} [tBid:Bax][Bax]}}{{k_{ -12} }} = K_{12} K_{11} [tBid][Bax][Bax], \\[0.15cm]
 [Cyto.c^* :Apaf:ATP] = & \frac{{k_{16} [Cyto.c^* ][Apaf][ATP]}}{{k_{ -16} }} = K_{16} [Cyto.c^* ][Apaf][ATP] , \\[0.15cm]
 [Cyto.c^* :Apaf:ATP:Casp9] = & \frac{{k_{17} [Cyto.c^* :Apaf:ATP][Casp9]}}{{k_{ -17} }} \\[0.15cm]
 = & K_{17} K_{16} [Cyto.c^* ][Apaf][ATP][Casp9], \\[0.15cm]
 [Cyto.c^* :Apaf:ATP:Casp9_2 ]= & \frac{{k_{18} [Cyto.c^* :Apaf:ATP:Casp9][Casp9]}}{{k_{ -18} }} \\[0.15cm]
 = & K_{18}K_{17} K_{16} [Cyto.c^* ][Apaf][ATP][Casp9]^2, \\[0.15cm]
 [Bcl_2 :Bax] = & \frac{{k_{24} [Bcl_2 ][Bax]}}{{k_{ -24} }} = K_{24} [Bcl_2 ][Bax], \\[0.15cm]
 [Bcl_2 :tBid] = & \frac{{k_{25} [Bcl_2 ][tBid]}}{{k_{ -25} }} = K_{25} [Bcl_2 ][tBid]
 \end{array}
 \end{equation}
with $K_i=k_i/k_{-i} $ for $i = 7, 9, 11, 12, 16, 17, 18, 24, 25$. Denote these relations by $X = \Psi(Y)$.

It is remarkable that the relations rely only on the 9 constants $K_i$, instead of the 18 constants $k_{\pm i}$. The latter are often not reliably known.

Substituting $X= \Psi(Y)$ into the second and third equations in (\ref{45}), we obtain
\[\left\{ \begin{array}{l}
  \displaystyle{ \frac{{d\tilde Y}}{{dt}} = Q_2(\Psi (Y), Y, Z)+ \Theta Q_1(\Psi (Y), Y, Z)}, \\[0.2cm]
    \displaystyle{ \frac{{dZ}}{{dt}} = Q_{3}(\Psi (Y),Y,Z)}
 \end{array} \right.\]
and thereby gain equations for $Y$:
$$
\displaystyle{ \frac{{dY}}{{dt}}= (I + \Theta\Psi(Y)_Y)^{-1}\frac{{d\tilde Y}}{{dt}}=(I + \Theta\Psi(Y)_Y)^{-1} (Q_2(\Psi(Y), Y, Z)+ \Theta Q_1(\Psi(Y), Y, Z))}.
$$
Consequently, the original system (\ref{31}) of 28 ODEs can be approximated by the following 19 ODEs
\begin{equation}\label{47}
\begin{array}{l}
  \displaystyle{ \frac{{dY}}{{dt}}=(I +\Theta\Psi(Y)_Y)^{-1}(Q_2(\Phi (Y),Y,Z) + \Theta Q_1(\Psi(Y),Y,Z))}, \\[0.2cm]
    \displaystyle{ \frac{{dZ}}{{dt}} = Q_{3}(\Psi (Y),Y,Z)}
\end{array}
\end{equation}
together with nine algebraic relations
$$
X = \Psi(Y)
$$
being detailed in (\ref{46}). Recall that $Y$ and $Z$ are defined in (\ref{41}). We call this new simplified model as ISS-2.

\section{Numerical simulations}

The purpose of this section is to show the reliability of our ISS-2 model by resorting to numerical simulations. Precisely, we compare the ISS-2 model (\ref{46})--(\ref{47}) with the entire ISS model (\ref{31}) and Okazaki et al.'s ISS skeleton model \cite{Okazaki} in several aspects, including the accuracy, M-D transition behavior and sensitivity. For the reader's convenience, the skeleton model is given in Table \ref{Tab:the ISS skeleton model}.
\begin{table}
\tiny
\vspace{0cm}
\caption{The ISS skeleton model due to Okazaki et al. (2008)
}
\medskip{}
\hspace{0cm}
\label{Tab:the ISS skeleton model}
\begin{tabular}{llll}
 \hline
Reaction & \space & Rate constant  \\
\hline
(S1) & $   Casp8_{_2 }^*  + Casp3  \autorightarrow {}{}   Casp8_{_2 }^* + Casp3^* $ & $6.25.00\times10^{-6}nM^{-1}s^{-1}$ \\[0.1cm]
(S2a) & $  Casp8_{_2 }^*  + Bid   \autorightarrow {}{}   Casp8_{_2 }^* + 0.0328tBid:Bax_2 $ & $ v_{S2a}=\frac{{k_a[Casp8^*_2][Bid]}}{{[Casp8^*_2]+K_a}}(k_a=0.1s^{-1},K_a=20nM)$ \\[0.1cm]
(S2b) & $ Cyto.c + tBid:Bax_2   \autorightarrow {}{}  0.867Cyto.c^*:Apaf:ATP + tBid:Bax_2 $ & $1\times10^{-3}nM^{-1}s^{-1}$ \\[0.1cm]
(S2c) & $ Cyto.c^* :Apaf:ATP + 2Casp9 \autorightarrow {}{}  Cyto.c^* :Apaf:ATP + Casp9 + Casp9^* $ & $1.46\times10^{-6}nM^{-1}s^{-1}$ \\[0.1cm]
(S2d) & $ Casp9^* + Casp3   \autorightarrow {}{}   Casp9^*  + Casp3^* $ & $1.96\times10^{-5}nM^{-1}s^{-1}$ \\[0.1cm]
\hline
\end{tabular}
\end{table}
The simulations were carried out with Matlab.

\subsection{Accuracy of the ISS-2 model}
\begin{figure}
\centering
    \subfigure[The curves of Casp3$^*$ as functions of $t$]{
        \label{Fig.Concentration of FSISS.1}
        \includegraphics[width=0.45\textwidth]{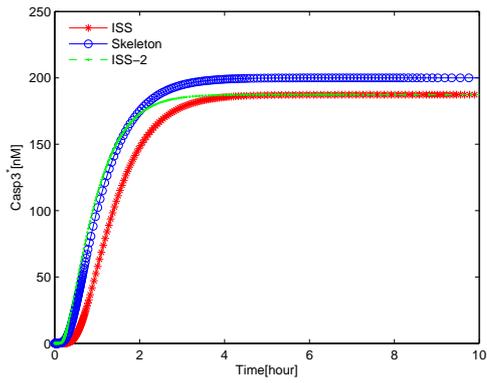}}
    \subfigure[The curves of Casp3 as functions of $t$]{
        \label{Fig.Concentration of FSISS.2}
        \includegraphics[width=0.45\textwidth]{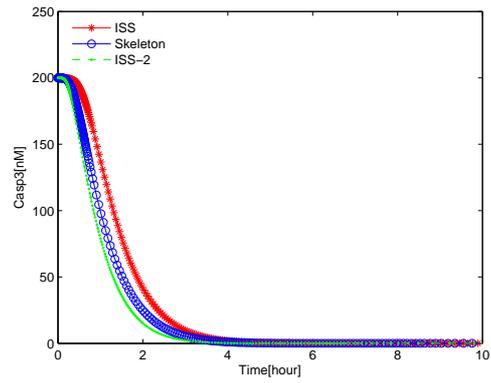}}
    \subfigure[The curves of Casp9$^*$ as functions of $t$]{
        \label{Fig.Concentration of FSISS.3}
        \includegraphics[width=0.45\textwidth]{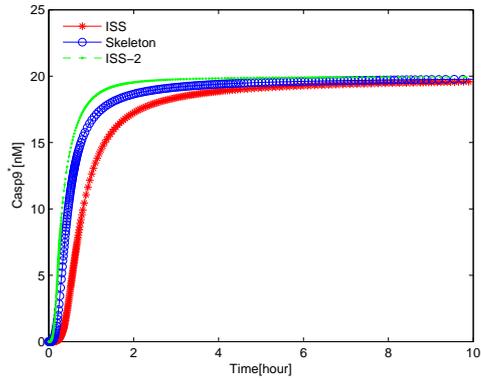}}
    \subfigure[The curves of Bid as functions of $t$]{
        \label{Fig.Concentration of FSISS.4}
        \includegraphics[width=0.45\textwidth]{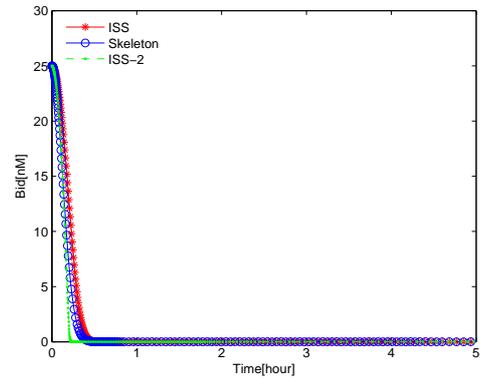}}
    \caption{Comparison of the ISS model (red asterisk), the skeleton model (blue open circle) and our ISS-2 model (green dot). The initial concentrations are from Table \ref{Tab:The initial concentrations of each species in ISS model}. (a):  the curves of Casp3$^*$ as functions of $t$. (b): the curves of Casp3 as functions of $t$. (c): the curves of Casp9$^*$ as functions of $t$. (d): the curves of Bid as functions of $t$.    }
    \label{Fig.Concentration of FSISS}
\end{figure}

We compute the concentration of each species as functions of time $t$ for the entire ISS model, the ISS skeleton model and our ISS-2 model, with initial concentrations from Table \ref{Tab:The initial concentrations of each species in ISS model}. Fig. \ref{Fig.Concentration of FSISS.1} displays three curves of Casp3$^*$ as functions of time $t$ corresponding to the three models. In this figure, the equilibrium value of Casp3$^*$---the indicator of apoptosis---
of the ISS-2 model is almost the same as that of the ISS model, whereas that of the skeleton model is slightly larger.
%So, we can say that our ISS-2 model matches the ISS model better than the ISS skeleton model does.
The equilibrium values of all other species for the ISS model and our ISS-2 model are also very close to each other. The curves of Casp3, Casp9$^*$ and Bid as functions of time $t$ are given in Fig. \ref{Fig.Concentration of FSISS.2}, Fig. \ref{Fig.Concentration of FSISS.3}, and Fig. \ref{Fig.Concentration of FSISS.4}, respectively. These numerical results show that our ISS-2 model is a reliable simplification of the entire ISS model.

%As a first step, we compute the concentration of each species as functions of time $t$, with the initial concentrations
%from Table \ref{Tab:The initial concentrations of each species in ISS model},
%by using the entire ISS model, the ISS skeleton model and our PSISS model. Fig. \ref{Fig.$Casp3^*$ comparison of ISS,
%Skeleton and PISS} displays the curves for Casp3$^*$ as functions of time $t$. From these figures, especially
%Fig. \ref{Fig.$Casp3^*$ comparison of ISS, Skeleton and PISS.2}, we see that the PSISS model and the ISS model give
%almost the same curves and equilibrium value for Casp3$^*$, while what the skeleton model yields are different. This
%conclusion is further supported by the curves for other species. For example, we see Fig. \ref{Fig.$Casp3$ and
%$Casp9^*$ comparison of ISS, Skeleton and PISS} for the curves of Casp3 and Casp9$^*$ as functions of time $t$.
%These numerical results show that our PSISS model is closer to the original ISS model than the skeleton model.

\subsection{M-D transition behavior}
The M-D transition behavior is explained at the end of Section 3. In \cite{Okazaki}, it was reported that initial concentrations of Casp9 also have considerable impacts to the transition behavior. In order to study this behavior, Okazaki et al. introduced two quantities $\gamma_D$ and $v^C_0$ in \cite{Okazaki}. The former was defined as the ratio of the net  production of Casp3$^*$ via the D-channel to its total production, while the latter
represents the critical value of $v_0$ (the generation rate for Casp8$_2^*$) corresponding to $\gamma_D=0.5$. Note that, at $\gamma_D=0.5$, the effect of the M-channel is same as that of the D-channel.

\begin{figure}
\centering
    \subfigure[M-D transition behaviors due to Casp8$_2^*$]{
    \label{Fig.M-D transition of FSISS.1}
        \includegraphics[width=0.45\textwidth]{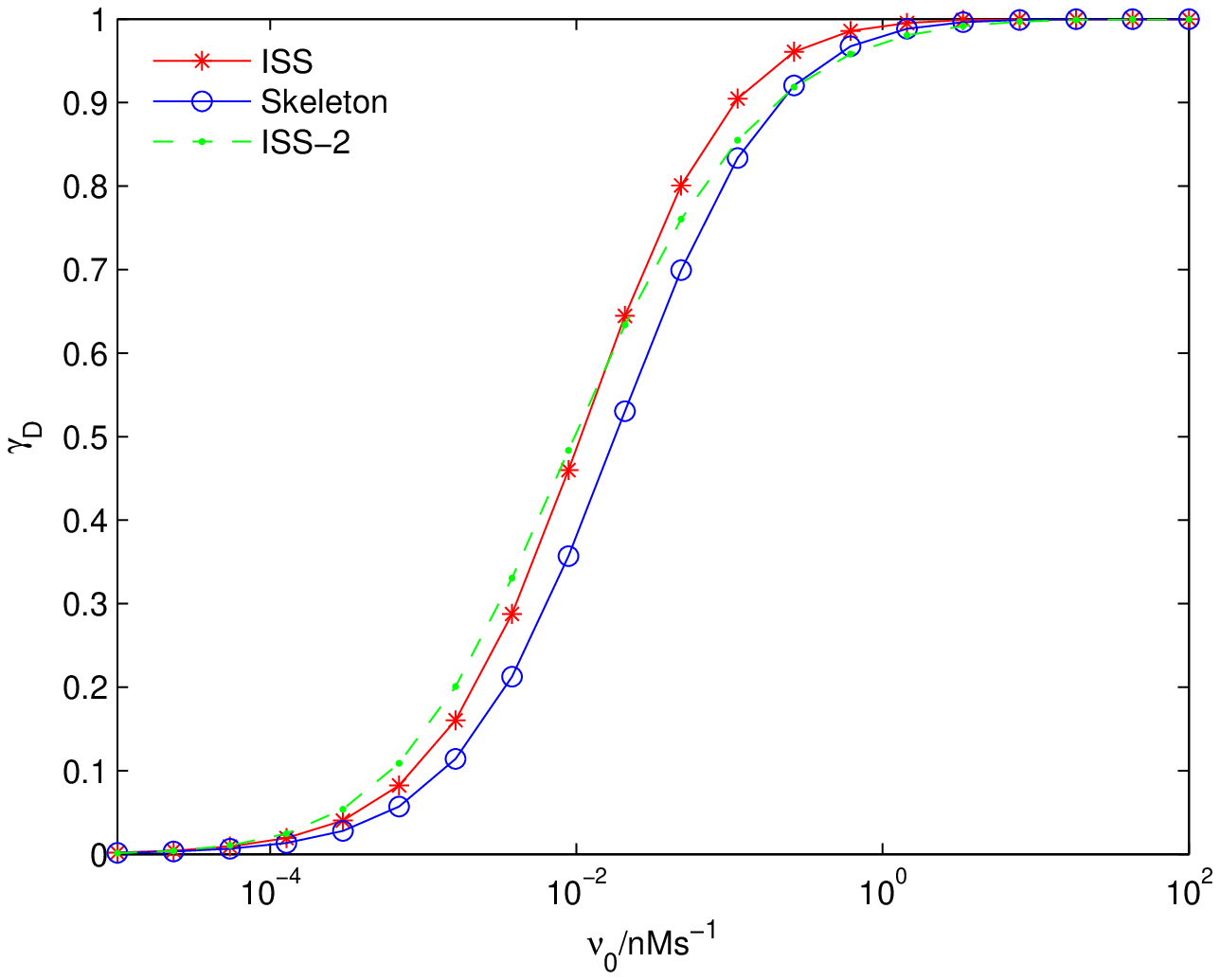}}
    \subfigure[M-D transition behaviors due to Casp9]{
    \label{Fig.M-D transition of FSISS.2}
        \includegraphics[width=0.45\textwidth]{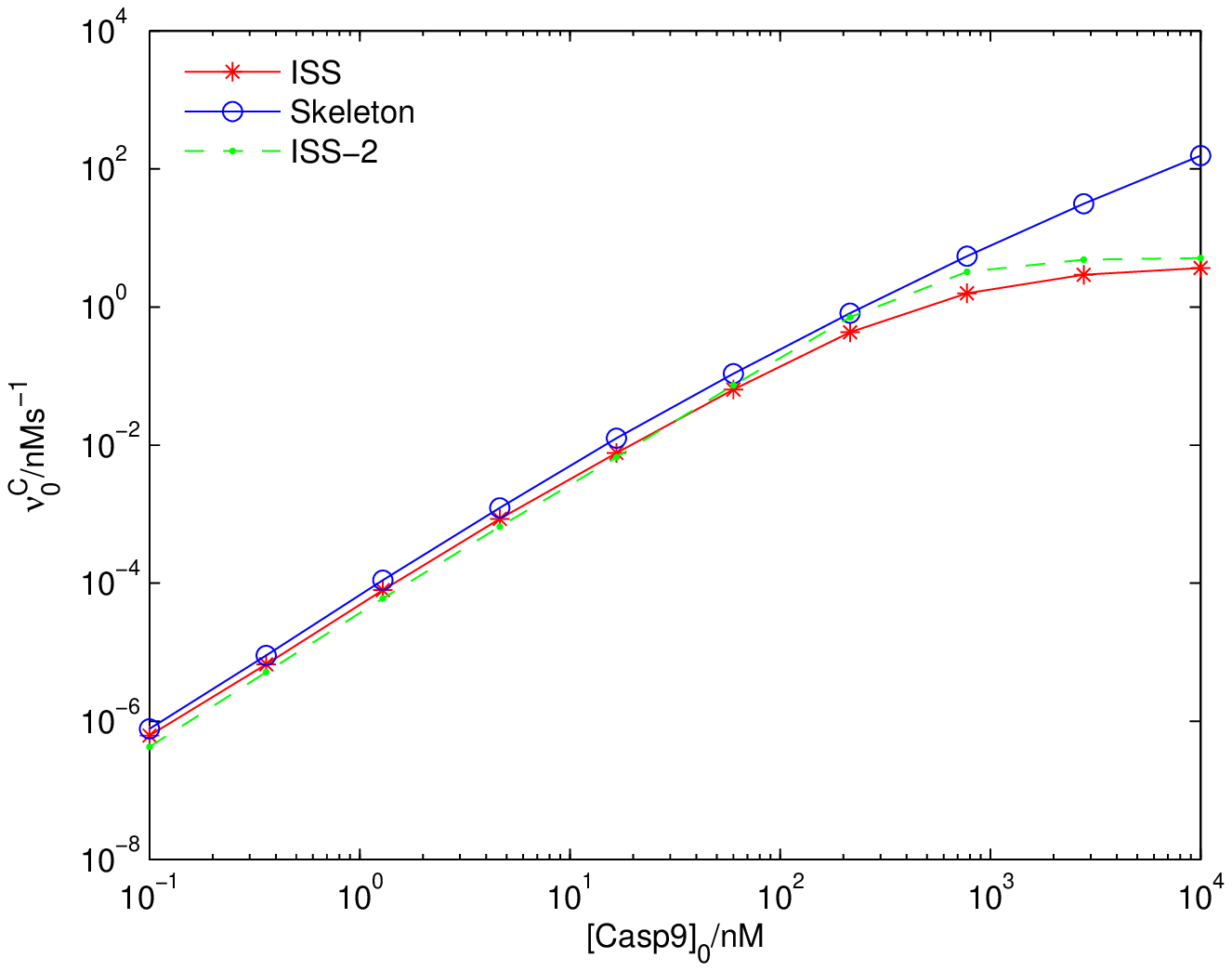}}
    \caption{Comparison of the ISS model (red asterisk), the skeleton model (blue open circle) and the ISS-2 model (green dot) for the M-D transition behavior. (a): M-D transition behavior due to Casp8$_2^*$. (b): M-D transition behavior due to Casp9. }
    \label{Fig.M-D transition of FSISS}
\end{figure}

As previously, we use the initial concentrations from Table \ref{Tab:The initial concentrations of each species in ISS model} and numerically solve the three models to obtain three curves of $\gamma_D$ as a function of $v_0$.
The results are shown in Fig. \ref{Fig.M-D transition of FSISS.1}. From this figure, we see that the curve given by  the ISS-2 model matches that by the ISS model quite well and is obviously better than that by the skeleton model.

The curves of $v^C_0$ as a function of the initial concentration of Casp9 are shown in Fig. \ref{Fig.M-D transition of FSISS.2}. From this figure we see that when the initial concentrations of Casp9 are small, the values of $v^C_0$ for the three models are almost the same. However, when the initial concentrations of Casp9 are large, the ISS skeleton model behaves quite different from the ISS model. But our ISS-2 model still matches the ISS model very well. All these indicate that our ISS-2 model can well describe the actual M-D transition behavior and are much better than the ISS skeleton model.

\subsection{Sensitivity Analysis}
\begin{figure}
\centering
    \subfigure[Half-time for activating Casp3 via the ISS model]{
    \label{Fig.Half-time for $Casp3^*$ activation.1}
        \includegraphics[width=0.45\textwidth]{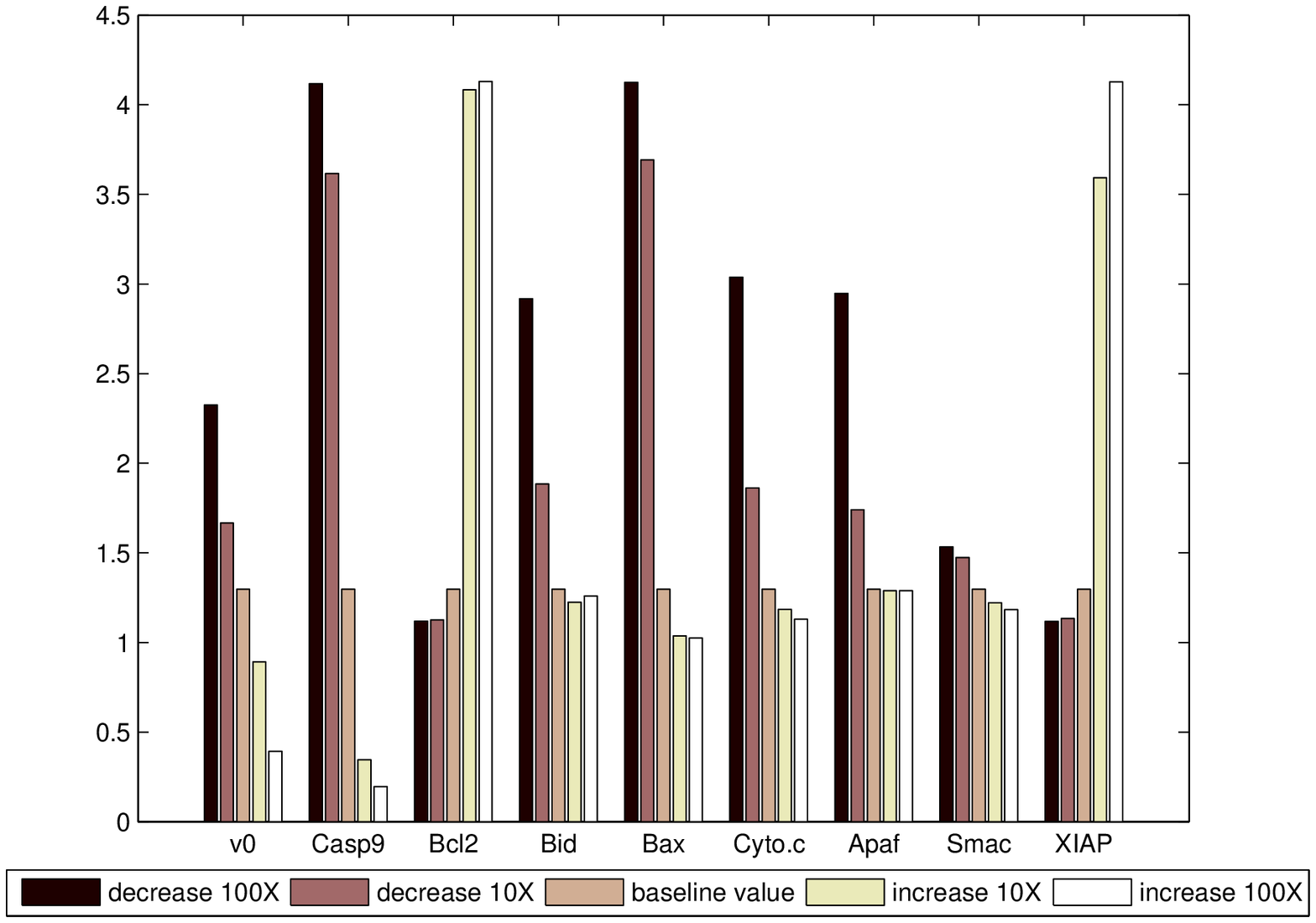}}
    \subfigure[Half-time for activating Casp3 via the skeleton model]{
    \label{Fig.Half-time for $Casp3^*$ activation.2}
        \includegraphics[width=0.45\textwidth]{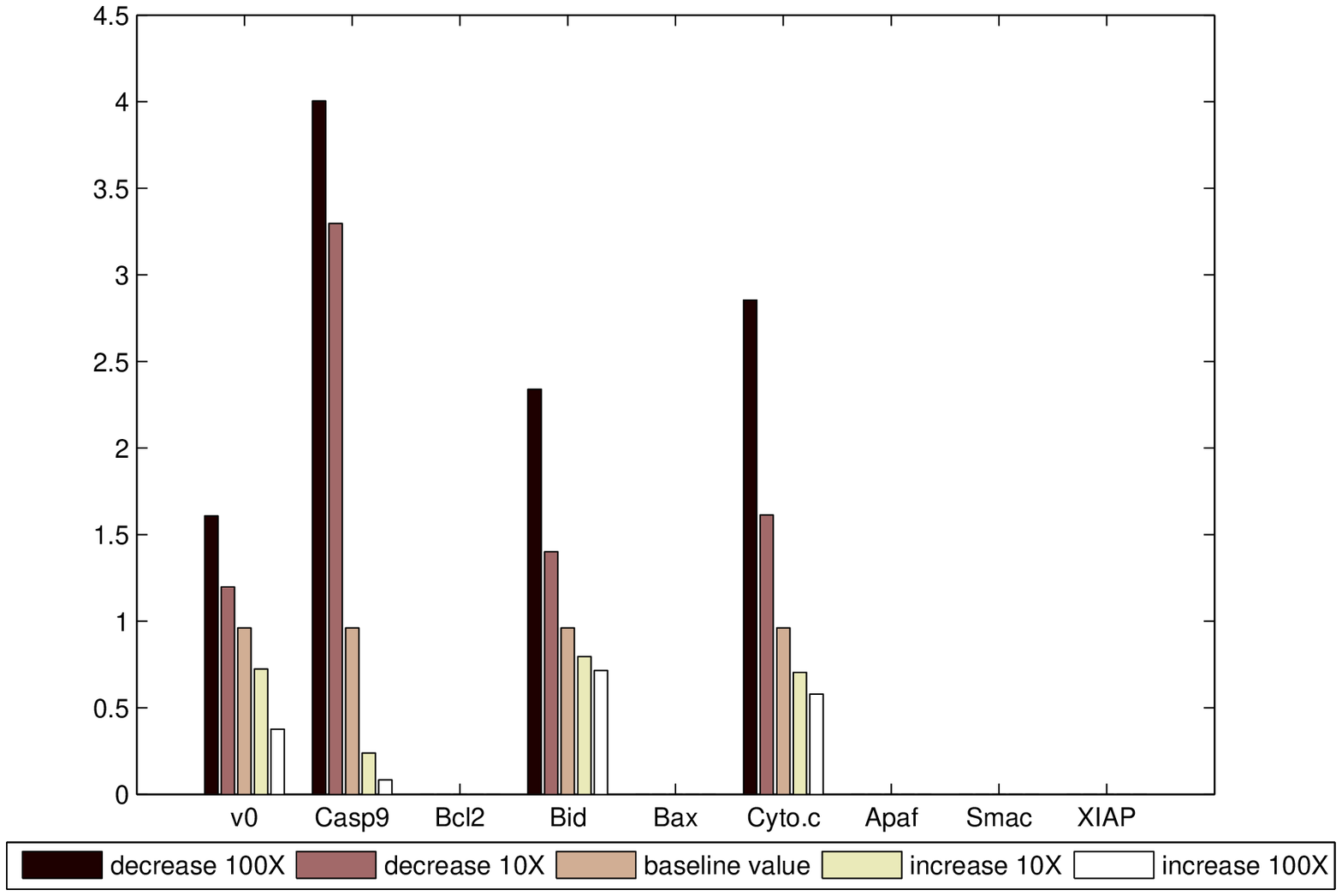}}
    \subfigure[Half-time for activating Casp3 via the ISS-2 model]{
    \label{Fig.Half-time for $Casp3^*$ activation.3}
        \includegraphics[width=0.45\textwidth]{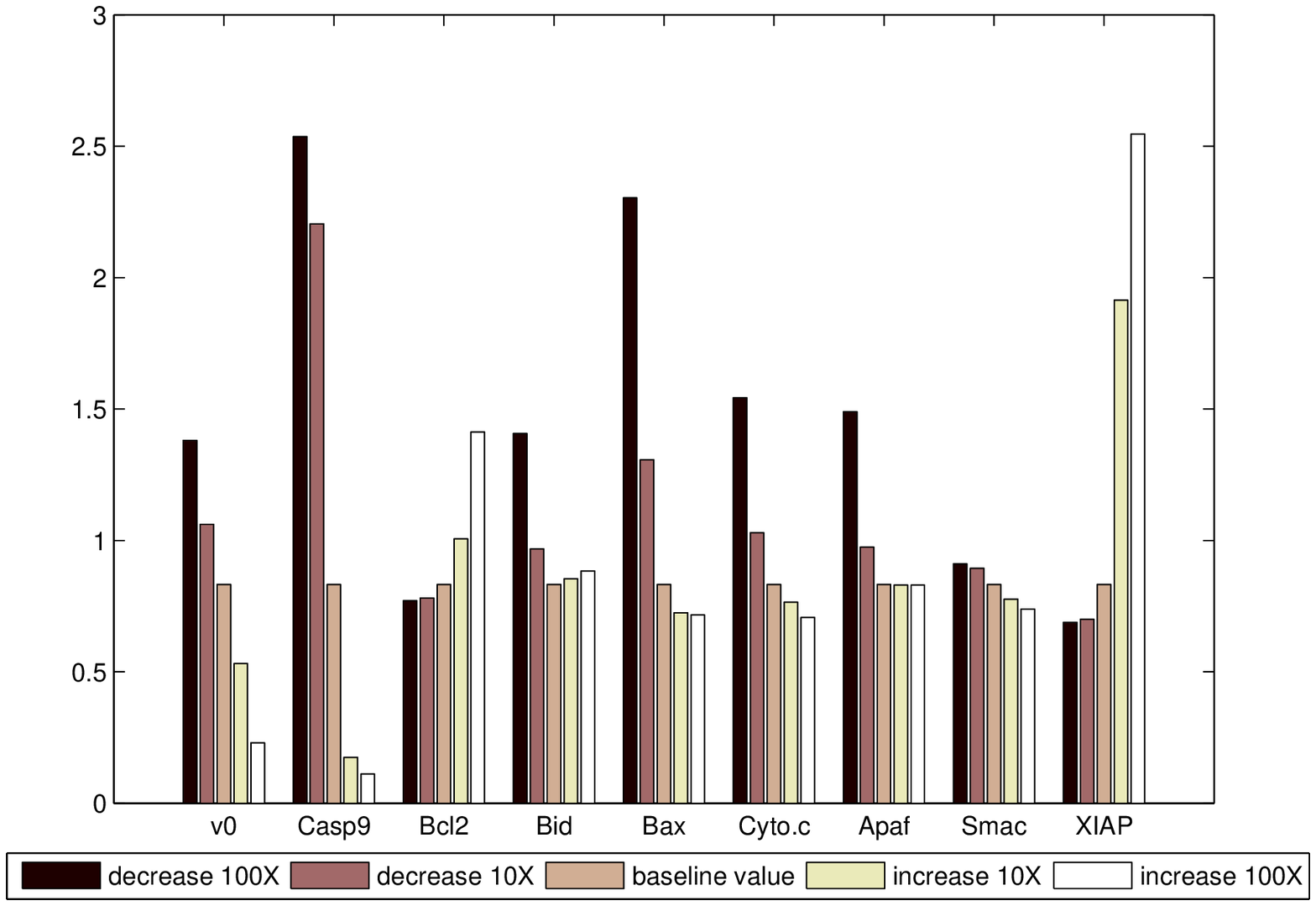}}
    \caption{Sensitivity analysis of the ISS model, the skeleton model and the ISS-2 model. The overexpression or knockdown level of each species is changed one or two orders of magnitude of the baseline values while the others are unchanged. }
    \label{Fig.Half-time for $Casp3^*$ activation.}
\end{figure}

Now we present some results on the sensitivity of our ISS-2 model. Because the half-time---the time for Casp3$^*$ to attain half of its equilibrium value---is an important quantity to characterize how fast a cell will die \cite{Hua},
we compute this quantity for the three models with initial data changed by
%increasing or decreasing the expression levels of each molecule.
one or two orders of magnitude higher and lower than the baseline values given in Table \ref{Tab:The initial concentrations of each species in ISS model}.
The numerical results are shown in Fig. \ref{Fig.Half-time for $Casp3^*$ activation.}.
%Let us mention that Hua et al. analyzed this sensitivity of their full model. The difference from \cite{Hua} is that
%our results are only for the molecules in the downstream process.

\begin{figure}%[h]
  \begin{center}
  \includegraphics[width=0.6\textwidth]{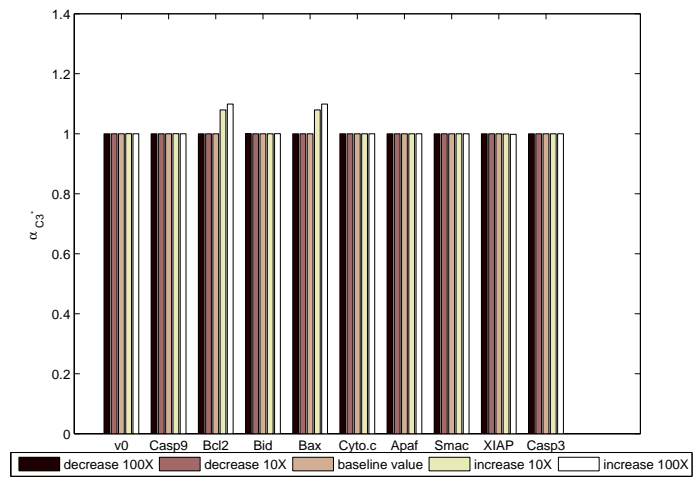}
  %\vspace{-.5em}
  \caption{The change of $\alpha _{C3^*}$ against initial concentrations. The overexpression or knockdown level of each species is changed one or two orders of magnitude of the baseline values while the others are unchanged.}
  \label{Fig:The matching of FSISS to ISS for different initial concentration.}
  %\vspace{1.2em}
  \end{center}
\end{figure}

From Fig. \ref{Fig.Half-time for $Casp3^*$ activation.} we see that, like the full apoptosis model due to Hua et al. \cite{Hua}, our ISS-2 model possesses the symmetrical or asymmetrical properties of varying each species to the outcome. The result by the ISS-2 model is very similar to that by the ISS model. A bit difference is that the half-time for the ISS-2 model is a little shorter than that for the ISS model, which is same as for the ISS skeleton model. This is expected because the assumption of fast reactions slightly speeds up the whole apoptotic process.

%The ISS sketeton model completely loses the information of this sensitivity of so many species.
%Overall, our ISS-2 model preserved this property of the ISS model better than the ISS sketeton model does.

To evaluate our simplified model, we follow our previous work \cite{HY} and introduce the quantity
$$
\alpha_{C3^*}=\frac{{equilibrium\ value\ of\ Casp3^*\ for\ ISS\mbox{-2}}}{{equilibrium\ value\ of\ Casp3^*\ for\ ISS}}
$$
to examine how different are the equilibrium values of Casp3$^*$ for the two models when changing initial concentrations of a certain species. When initial concentrations of some species are changed, $\alpha_{C3^*}$ will likely change too. For a good simplified model, such a quantity should be close to one.

%to examine the sensitivity of Casp3$^*$ concentration match of our ISS-2 model with the ISS model at equilibrium to
%different initial concentrations.
%to examine the equilibrium values of Casp3$^*$ by changing the initial concentrations of a certain species.

We compute $\alpha_{C3^*}$ for changing initial concentrations of each species, including Casp3$^*$, by one or two orders of magnitude higher and lower than the baseline values as before. The result is given in Fig. \ref{Fig:The matching of FSISS to ISS for different initial concentration.}. This result illustrates that $\alpha_{C3^*}$ is insensitive to most of initial concentration changes, except a little sensitivity for Bcl2 and Bax.
%In total, $\alpha_{C3^*}$ is almost free from the initial concentration changes.
In conclusion, our ISS-2 model well retains the main features of the ISS model and therefore can be viewed as a reliable simplification to the original ISS model.

\section{Summary}

In this paper, we develop a general framework of the PEA method together with two conditions, under which
the method can be justified rigorously. These conditions are the fastness assumption and the principle of detailed balance for fast reactions as a whole. Under these conditions, we simplify a general system of chemical reactions governed by the law of mass action. This simplification clearly has a solid mathematical basis.

Then we follow the general framework and study the ISS (intracellular--signaling subsystem) model as the downstream process of the Fas-signaling pathway model proposed by Hua et al. (2005) for human Jurkat T cells. Because the framework has a solid mathematical basis, it can be used as a tool to determine whether or not a reaction is relatively fast. After many attempts via numerical tests, we found that nine of reactions in the ISS model can be well regarded as fast.

Knowing that the nine reactions are faster than others, we use the justified PEA method and simplify the ISS model to derive a so-called ISS-2 model. It is remarkable that the reversible reactions in apoptosis obey the principle of detailed balance naturally. With numerical simulations, we compare the ISS-2 model with the ISS model as well as Okazaki's ISS skeleton model in several aspects, including the accuracy, M-D transition behavior and sensitivity analysis. All the simulations show that the ISS-2 model is reliable. In particular, the new model can very well capture the M-D transition behavior of the ISS model at large initial concentrations of Casp9 and therefore improves Okazaki's ISS skeleton model considerably (see Fig. \ref{Fig.M-D transition of FSISS.2}).

At present, we are trying to simplify the upstream process with the justified PEA method. In the future, we will also try to simplify the whole process by correctly combining the PEA method and the QSSA method.

\section*{Acknowledgments}

This work was supported by the National Natural Science Foundation of China (NSFC 10971113) and by Specialized Research Fund for the Doctoral Program of Higher Education (Grant No. 20100002110085).

\end{spacing}
\end{CJK*}
\end{document}